\documentclass{article}
\usepackage[utf8]{inputenc}
\usepackage{amsfonts,amsthm,amsmath,amssymb, verbatim}
\usepackage{array}
\usepackage{epsfig}
\usepackage{fullpage}
\usepackage{xcolor}
\usepackage{graphicx}
\usepackage{caption}
\usepackage{wrapfig}
\usepackage{lscape}
\usepackage{rotating}
\usepackage{epstopdf}
\usepackage[font=small,labelfont=bf]{caption} 
\newtheorem{theorem}{Theorem}
\usepackage{hyperref}

\usepackage[
backend=bibtex,
style=alphabetic,
sorting=ynt
]{biblatex}
\def\defn#1{\textbf{\textit{\boldmath #1}}}

\title{ZHED is NP-complete}
\author{%
  Sagnik Saha%
    \thanks{Work done while at MIT.}
\and
  Erik D. Demaine%
    \thanks{MIT Computer Science and Artificial Intelligence Laboratory,
      32 Vassar St., Cambridge, MA 02139, USA, \protect\url{edemaine@mit.edu}}
}
\date{}

\begin{document}

\maketitle

\begin{abstract}
  We prove that the 2017 puzzle game ZHED is NP-complete, even with just
  $1$ tiles.
  Such a puzzle is defined by a set of unit-square $1$ tiles in a
  square grid, and a target square of the grid.
  A move consists of selecting an unselected $1$ tile and then filling
  the next unfilled square in a chosen direction from that tile
  (similar to Tipover and Cross Purposes).
  We prove NP-completeness of deciding whether the target square can be filled,
  by a reduction from rectilinear planar monotone 3SAT.
\end{abstract}

\section{Introduction}

\begin{figure}[h!]
    \centering
    \includegraphics[scale=0.5]{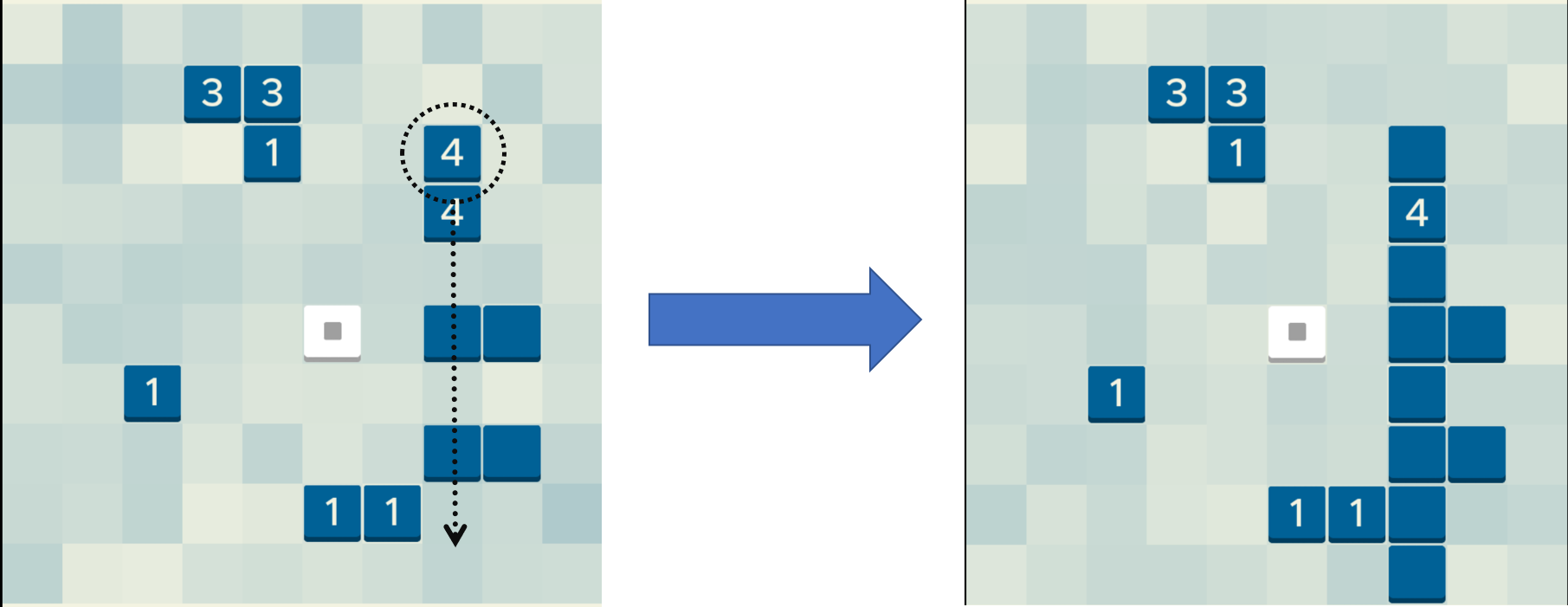}
    \caption{A typical move in ZHED. Tiles are drawn in dark blue.
      The target square has a small square in the middle.}
    \label{fig:move}
\end{figure}

ZHED \cite{zhed}
is a puzzle game available for Android, iOS, Switch, and Steam,
first released in 2017.
An instance of this puzzle is played on a $n\times n$ square grid;
refer to Figure~\ref{fig:move} (left).
One of the squares is designated as the \defn{target}.
Several of the other squares are \defn{filled} by \defn{tiles}.
Each initial tile has an integer \defn{number}
between $1$ and $n-1$ written on it.

In each move, the player selects one of the remaining numbered tiles
and a direction (up, down, left, or right);
refer to Figure~\ref{fig:move}.
If the tile was numbered $k$, the move replaces the tile with a blank tile
(keeping the square filled, but removing the number)
and fills the $k$ closest unfilled squares
in the specified direction with blank tiles.
The objective of the game is to fill the target square.

The number of moves in such a puzzle is bounded by the number of numbered
tiles, so the puzzle is in NP.
In this paper, we prove that the puzzle is in fact NP-complete,
even when all tiles are numbered $1$.

The mechanic of ZHED moves is similar to the idea of a tower of specified
height that falls over in a specified direction.
This ``falling tower'' mechanic is present in the puzzle board game Tipover,
which is NP-complete \cite{tipover},
and the 2-player board game Cross Purposes,
which is PSPACE-complete \cite{crosspurposes}.
The key distinction is that both Tipover and Cross Purposes prevent a tower
from falling on top of occupied/filled squares.

\section{Rectilinear Planar Monotone 3SAT}

\begin{figure}[h!]
    \centering
    \includegraphics[width=0.9\linewidth]{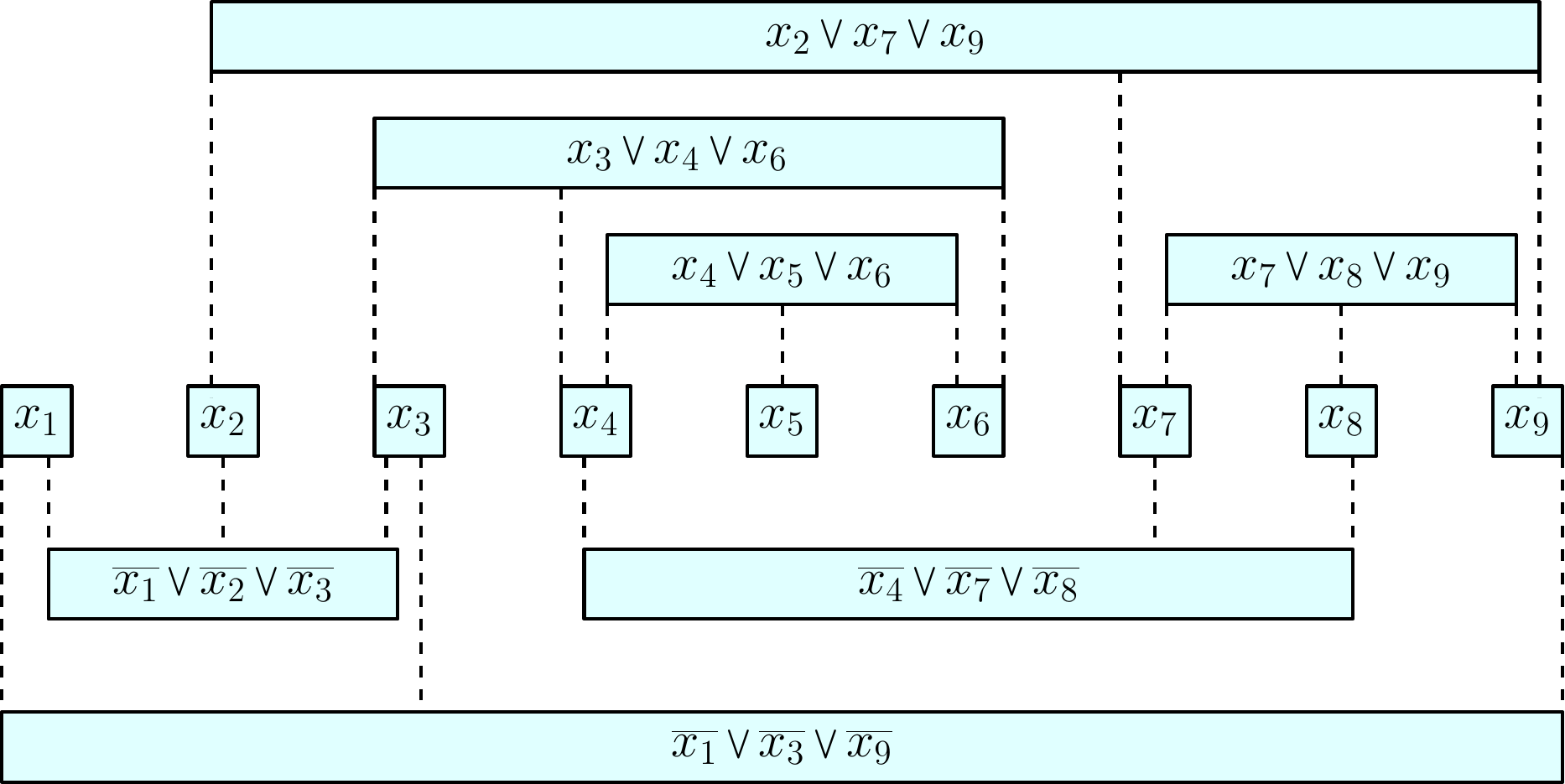}
    \caption{A sample instance of Rectilinear Planar Monotone 3SAT (RPM-3SAT).
      This planar embedding represents the conjunctive-normal-form formula
      $
      (x_2 \vee x_7 \vee x_9) \wedge
      (x_3 \vee x_4 \vee x_6) \wedge
      (x_4 \vee x_5 \vee x_6) \wedge
      (x_7 \vee x_8 \vee x_9) \wedge
      (\overline{x_1} \vee \overline{x_2} \vee \overline{x_3}) \wedge
      (\overline{x_4} \vee \overline{x_7} \vee \overline{x_8}) \wedge
      (\overline{x_1} \vee \overline{x_3} \vee \overline{x_9})
      $.}
    \label{fig:rpm-3sat}
\end{figure}

Our NP-hardness reduction is from the known NP-complete problem
Rectilinear Planar Monotone 3SAT (henceforth called \defn{RPM-3SAT})
\cite{rmp3sat};
refer to Figure~\ref{fig:rpm-3sat}.
In this problem, we are given and a Boolean formula $\mathcal{F}$
over $n$ Boolean variables $x_1, x_2, \dots, x_n$.
Formula $\mathcal F$ is in conjunctive normal form, so
the formula $\mathcal{F}$ is the logical AND ($\wedge$) of $m$ \defn{clauses}.
Each clause in $\mathcal{F}$ is a logical OR ($\vee$) of at most three
\defn{literals}, and the literals in each clause are either all positive,
consisting of unnegated variables of the form $x_i$, or all negative,
consisting of negated variables of the form~$\overline{x_i}$.
(The last is the ``monotonicity'' property.)
We are also given a planar embedding of the variable--clause incidence graph
connecting each variable to the clauses containing them,
which satisfies the following three conditions:
\begin{enumerate}
    \item All variables and clauses are grid-aligned rectangles of height $1$.
    \item All of the variables lie along a single horizontal line.
    \item All edges lie along vertical lines.
    \item All positive clauses lie above the line of the variables.
    \item All negative clauses lie below the line of the variables.
\end{enumerate}
The goal in the RPM-3SAT problem is to determine whether there exists a
value assignment to the $n$ Boolean variables that satisfies $\mathcal{F}$.
It is known to be NP-complete \cite{rmp3sat}.

\section{Reduction: Puzzle Construction}
\label{sec:reduction}

We now show how to build a ZHED puzzle that represents an arbitrary instance of RPM-3SAT, in such a way that the puzzle we construct will be solvable if and only if the original problem is satisfiable
(as proved in Section~\ref{sec:correctness}.
We start with a description of the various gadgets in our construction,
and then describe how they are combined together.
In our construction, we only need to use tiles with the number $1$,
so in the figures we omit the number on tiles.

\subsection{Threshold Gadget: Wires, AND, OR}

\begin{figure}[h!]
    \centering
    \includegraphics[width=\textwidth,page=2]{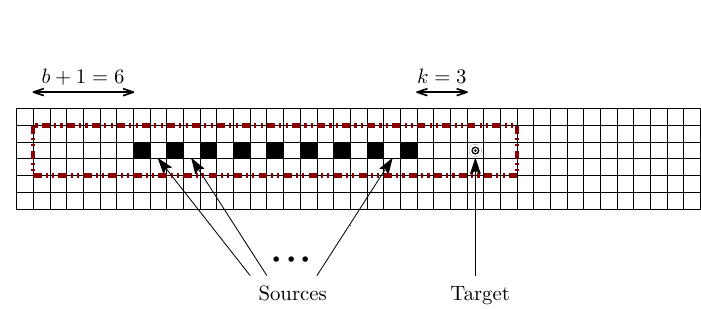}
    \caption{Threshold gadget. We assume that at most 5 of the source squares can be filled by outside tiles, and this gadget activates as long as at least 3 of those squares are filled before the tiles in this gadget are expanded rightwards.}
    \label{threshold_pic}
\end{figure}

Our main workhorse is the \defn{threshold gadget},
which consists of a line of tiles on alternate squares,
as shown in Figure~\ref{threshold_pic}.
The empty squares between the tiles each function as a \defn{source}.
The threshold gadget is parameterized by a nonnegative integer $k$,
and we call the square at a distance of $k+1$ from the last tile in the gadget
(on the same line) the \defn{target}.

\paragraph{Threshold property.}
This gadget has the property that the target square can be filled
if and only if at least $k$ of the source squares were already
filled by tiles from other gadgets.
The intended activation is to expand each tile in the gadget toward the target,
in decreasing order of distance.
Without any sources already filled, this will reach only one square from the
last tile in the gadget.
If, however, $j$ sources were already filled, then it will reach distance $j+1$.
Thus, the target will get filled if and only if $j \geq k$.

\paragraph{Wire, turn, AND, OR.}
This gadget is versatile, and forms the basis of several other gadgets.
If we connect only one source to other gadgets and set $k=1$,
then we get simple \defn{wire gadget} which propagate a signal
(represented by a square being filled) from that source to the target.
We can \defn{turn} the wire by connecting the target of one wire at
a source of another orthogonal wire.
On the other hand, if $m > 1$ sources are attached to other gadgets
(overlapping each source with the target of the other gadget,
representing $m$ input signals), then we obtain an \defn{OR gadget}
by setting $k=1$, and we obtain an \defn{AND gadget} by setting $k=m$.

\paragraph{Bounding box.}
Because of the spacing between the tiles, a threshold gadget can only
possibly fill squares on that line within a distance of $b+1$
from one end of the gadget, where $b$ is the maximum number of sources
that can be filled by other gadgets.
Because each tile has the number $1$, the gadget can also affect at most
the two adjacent lines of squares.
Therefore, the dotted rectangle in Figure~\ref{threshold_pic} serves as a
\defn{bounding box} of the gadget.
No sequence of moves in the overall puzzle can lead to a tile
in the threshold gadget directly filling a square outside its bounding box.

\paragraph{Chaining threshold gadgets.}

\begin{figure}[h!]
    \centering
    \includegraphics[width=\textwidth,page=1]{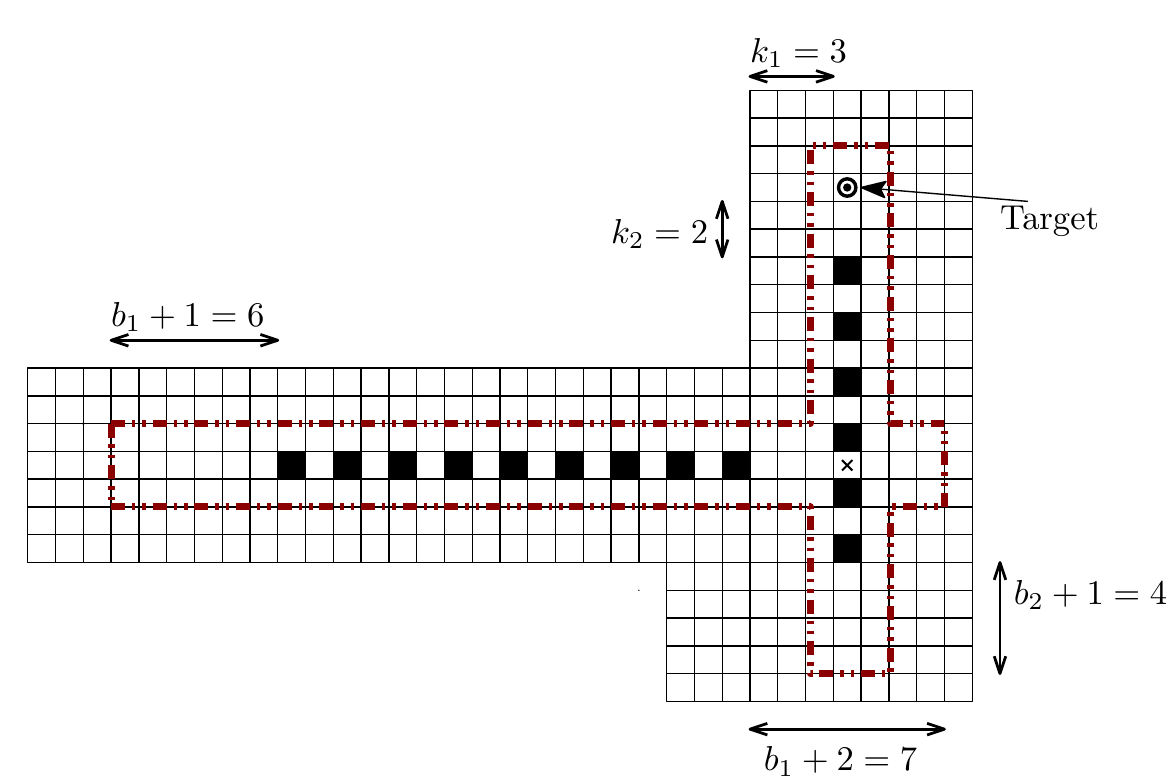}
    \caption{Chaining together two threshold gadgets (with parameters $k_1$ and $k_2$ respectively). The target of the horizontal threshold gadget (marked by $\times$) acts as a source for the vertical threshold gadget.}
    \label{chaining_threshold_pic}
\end{figure}

As eluded to above, we can chain threshold gadgets by making the target of one
a source of the next, as shown in Figure~\ref{chaining_threshold_pic}.
This allows us to aggregate signals and turn wires,
as chaining necessarily introduces a right-angle turn.
During activation, we simply expand the first gadgets' tiles first,
which (maybe) fills its target square,
and then expand the tiles of the second gadget.

This interaction is one of only two ways different gadgets are allowed
to interact with each other in our construction.
The first gadget's bounding box grows by $1$ unit in the direction it
propagates, because the second gadget may be expanded before it.
The composite bounding box is simply the union of the two gadgets'
individual bounding boxes.

\subsection{Shift Gadget}

\begin{figure}[h!]
    \centering
    \includegraphics[width=\textwidth,page=7]{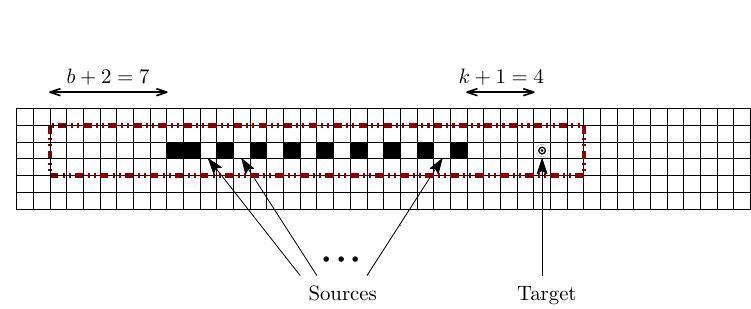}
    \caption{Shift gadget. The threshold gadget from Figure~\ref{threshold_pic} is shifted by one square to the right.}
    \label{shift_pic}
\end{figure}

To resolve parity issues in our construction, we sometimes need to shift
threshold gadgets by $1$ square. To do so, we simply add an extra tile at the
beginning of the gadget, adjacent to the old starting tile, as in
Figure~\ref{shift_pic}.
To activate this gadget, we expand all tiles toward the target,
in decreasing order by distance (as in the threshold gadget).

\paragraph{Bounding box.}
As a result of the extra tile, the bounding box of the original threshold gadget
expands by $1$ square in the forward direction, and $2$ squares in the backward
direction (counting the new tile itself, as well as the extra square it can
potentially fill).
We move the target square one square forward,
and everything else remains unchanged.

\subsection{Variable Gadget}

\begin{figure}[h!]
    \centering
    \includegraphics[width=\textwidth,page=4]{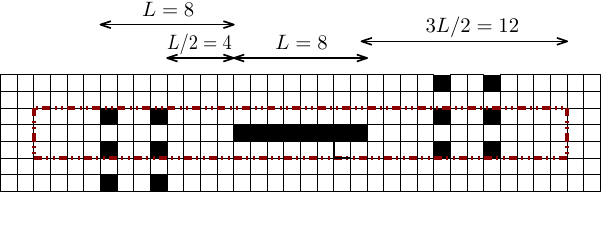}
    \caption{Variable gadget with $L = 8$ tiles, along with four threshold gadgets ``reading'' the value of this variable. The two threshold gadgets right of the variable can get one of their source squares filled if the variable tiles are all expanded right, setting it to true. The two gadgets on the left can be activated when the variable is set to false.}
    \label{variable_pic}
\end{figure}

A \defn{variable gadget} consists of a large even number $L$ of tiles
consecutive on a horizontal row; refer to Figure~\ref{variable_pic}.
These tiles are all meant to be expanded in the same direction,
either left or right.
In our construction, expanding all tiles left corresponds to setting the
variable to \defn{false}, and expanding all tiles to the right
corresponds to setting the variable to \defn{true}. 

Activating a variable gadget as intended fills $L$ squares in one direction.
We attach vertical wires that propagate the filling of certain squares
to the left or right of the tiles (to clause gadgets),
thus ``reading'' the value of the variable.
All such wires are at a distance between $L/2 +1$ and $L$
from the variable gadget.
This choice ensures that an unintended activation of the gadget,
which expands some tiles left and other tiles right,
cannot activate wires on both sides of the variable.
For example, if $x$ of tiles are expanded left and $y$ are expanded right,
then at least one of $x$ and $y$ will be less than $L/2 + 1$,
so no wire in that direction will be activated.

We choose the value of $L$ based on the number of vertical wires needed.
In the rest of the construction described below, we specify the number of wires
needed for each clause gadget, and the gaps needed between consecutive wire.
Any value of $L$ large enough so that all of those wires fit within
$L/2 - 1$ columns suffices.

\paragraph{Bounding box.}
The bounding box of the variable gadget is again only $3$ rows high
because all the tiles have number $1$.
There are less than $L/2$ vertical wires on either side
within the reach of the tiles in this gadget,
so we obtain a upper bound on the bounding box of the variable gadget
by supposing that all of those wires were already expanded
before any tile in this gadget is expanded.
Thus the bounding box of a variable gadget spans at most $L + L/2 = 3L/2$
columns outside the tiles in each direction,
as shown in Figure~\ref{variable_pic}.

\subsection{Clause Gadget}

\begin{figure}[h!]
    \centering
    \includegraphics[width=\textwidth,page=5]{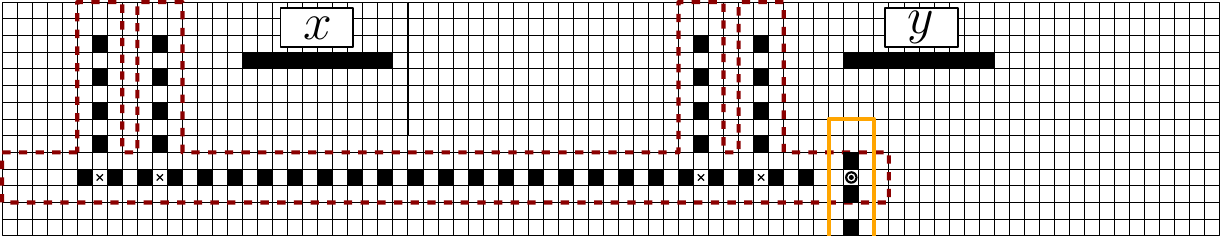}
    \caption{Clause gadget for $(\overline{x}\vee\overline{y})$. We use thick wires, each made of 2 individual wires, to connect the clause with the variables. We refer to the combination of the thick wires and the horizontal OR gadget as the clause gadget. The vertical wire chained from the OR gadget (with the orange bounding box) is a clause propagator that uses the target square of this clause gadget as a source, and propagates the signal elsewhere.}
    \label{clause_pic}
\end{figure}

The threshold gadget enables us to build a \defn{clause gadget};
refer to Figure~\ref{clause_pic}.
Because our 3SAT instance is monotone,
each clause uses only positive or only negative literals.
For positive clauses, we create a horizontal threshold gadget
going rightwards above the variables;
and for negative clauses, we create the threshold gadget below the variables.
Then we connect the clause to corresponding variables
(on the left side of the variable for positive clauses, and
on the right side of the variable for negative clauses)
via vertical ``thick wires''.
A \defn{thick wire} is a group of $g$ parallel wires connecting
the same gadgets, each separated $4$ units apart from each other;
thus, a single input actually advances the reach of the gadget by~$g$.
For the threshold gadget to function as an OR of the literals,
we set its $k$ to $g$ (instead of~$1$).
Thus the target square for the clause gadget is $g+1$ distance away
from the last tile in the gadget.

\paragraph{Bounding boxes.}
The clause gadget consists mainly of a threshold gadget
which can have $3g$ filled sources, so it already has a defined bounding box.
By our choice of separation, the bounding boxes of the individual wires
comprising each thick wire do not overlap.

\subsection{Crossover Gadget}

\begin{figure}[h!]
    \centering
    \includegraphics[width=\textwidth,page=3]{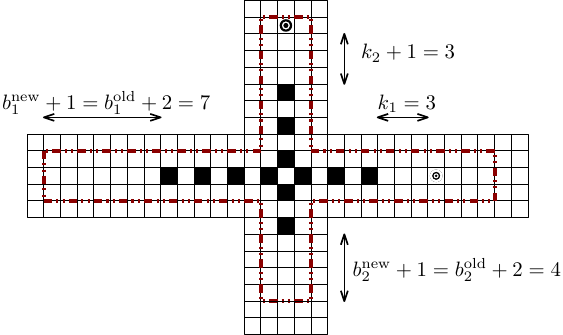}
    \caption{Crossover gadget. The horizontal threshold gadget is meant to be expanded first, so its target remains unchanged. The vertical threshold gadget is to be activated second, when the intersection square is already filled in. This pushes the target by one square.}
    \label{crossover_pic}
\end{figure}

We use a \defn{crossover gadget} when we need a vertical threshold gadget to
intersect the horizontal threshold gadget belonging to a clause.
As shown in Figure~\ref{crossover_pic}, we simply position the two
threshold gadgets so that the intersection square is blank
(with four tiles surrounding that intersection square like a plus sign).
Next, we increase the $k$ of the vertical threshold gadget by $1$.
Essentially, we assume that the horizontal gadget will be activated
before the vertical one, and we compensate for the extra filled source
square for the vertical gadget by moving its target square one unit farther.
If the player violates this assumption, the horizontal gadget will effectively
have its $k$ off by $1$, an error we will tolerate using thick wires
(see below).

\paragraph{Bounding box.}
The bounding box of the crossover gadget is easy to calculate
because there are only four tiles involved.
At the point of intersection, a $3 \times 3$ square centered at the
intersection point suffices.
The bounding boxes of both gadgets involved get longer in their long direction
by $2$ units ($1$ unit on each end)
because of the potential extra source square that may be filled.

\subsection{Putting It All Together}

\begin{sidewaysfigure}
\centering
    \includegraphics[angle=270,width=\textwidth,page=8]{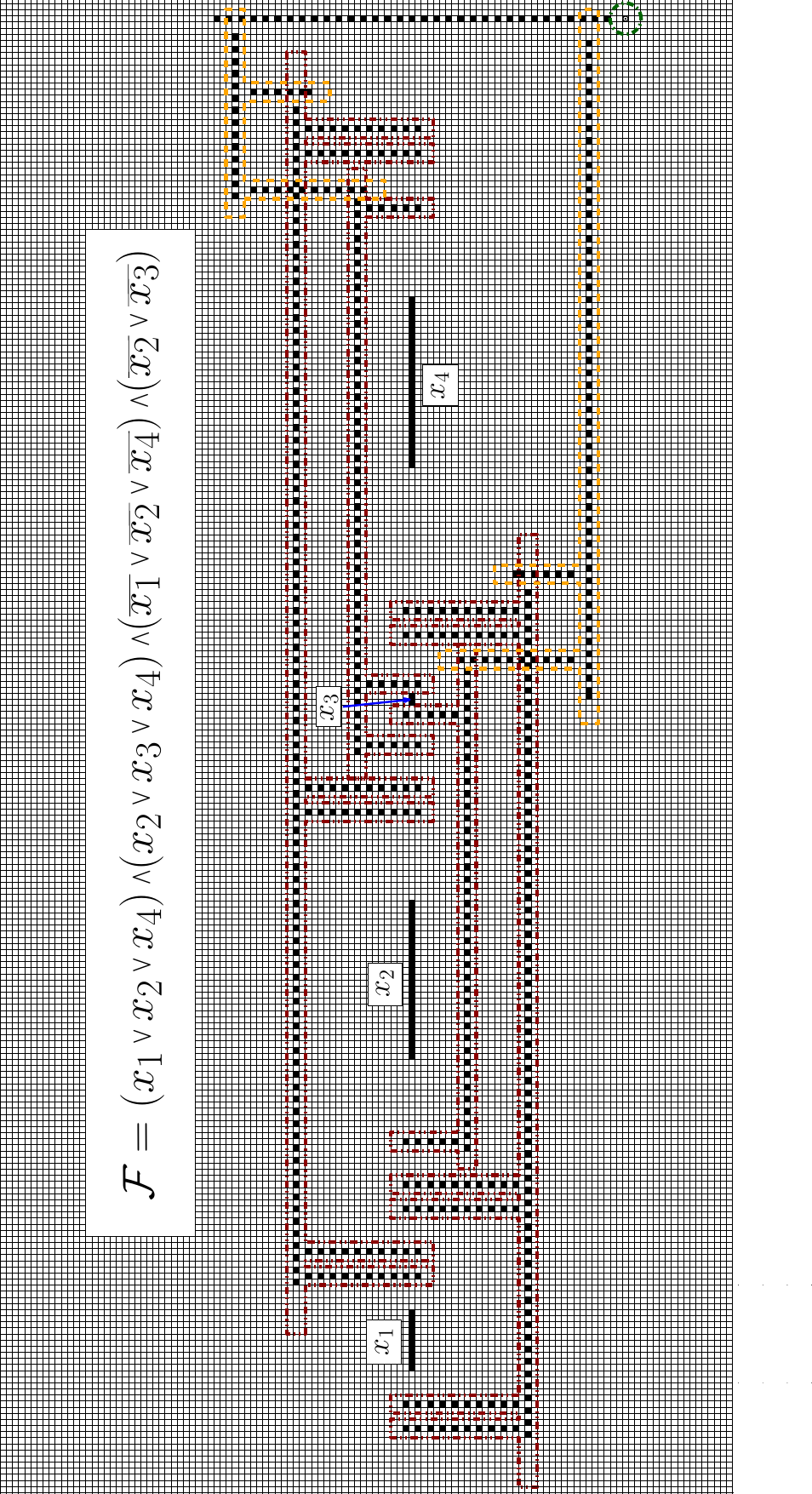}
    \caption{The entire puzzle board for a sample formula. The bounding boxes for the clause gadgets are marked in red. The bounding boxes of the clause propagators together with the upper and lower AND gadgets are colored orange. The final target square is at the lower-right of the board, marked by a green circle.
    Note that different variables have different sizes $L$ depending on how many wires connect it to clause gadgets: $v_3$ uses only $2$ tiles while $v_4$ uses $28$. Similarly, the first and third clauses use thick wires while the other two use single wires to connect with variables. We use shift gadgets for the lower AND gadget and for a couple of clause propagators.}
    \label{entire_pic}
\end{sidewaysfigure}

To construct the overall puzzle, we transform the planar embedding of
the provided RPM-3SAT instance; refer to Figure~\ref{entire_pic}.
All the variable gadgets go on the central horizontal row,
spaced far enough apart that none of their bounding boxes overlap.
We then draw the clause gadgets according to the provided clause rectangles
(above and below the central row for positive and negative clauses
respectively).
We space out the clauses more than the original drawing in order to leave room
for the thick-wire connections between variables and clauses.
Because we started with a planar monotone drawing,
we obtain a noncrossing drawing.

Next we add vertical wires
(henceforth referred to as \defn{clause propagators})
to transport the output signals from all of the target squares
of the positive clauses up.
We choose a row above the highest positive clause,
and place an AND gadget horizontally on that row,
with the target of all positive clause propagators coinciding with
sources of the AND gadget.
We perform a symmetric construction below the negative clauses,
resulting in an upper and a lower AND gadget.

Each clause propagator may intersect with the horizontal threshold gadgets
from some of the other clauses (nested above this one).
We use a crossover gadget for each such point of intersection.
For each clause, we count the number $x$ of crossovers in its OR gadget,
and set the thickness $g$ of the thick wires attached to the clause gadget
to be strictly larger: $g > x$.
This choice ensures that all the propagators intersecting a clause combined
have less influence on the clause threshold gadget than a single thick wire,
so no clause target can be satisfied by the propagator intersections alone.

We design the upper and lower AND gadgets to both go rightwards and
have their targets on the same column,
which is farther right than the bounding box of every other gadget.
We combine these two signals using a vertical AND gadget,
with a target square near the bottom right of the board.
This square is the actual \defn{target}
for the entire puzzle in our construction.

During this construction, we space the gadgets out sufficiently so that
no two bounding boxes collide except as part of an intended interaction
(through crossover gadgets or chaining threshold gadgets).
The thickness $g$ of each clause gadget's thick wires are determined
at this stage, and the size $L$ of each variable gadget
then depends on the total number of individual wires using it.
Because all of our gadgets have bounding-box dimensions of polynomial size,
the overall board remains polynomial sized in the input parameters.

The parity issue in the above construction is that the threshold gadget
has a repeating unit of size $2$, and hence we need to worry about
parity in lining everything up.
This issue may cause problems when we want multiple threshold gadget targets
to be on the same line for a chained gadget.
Parity issues may also become significant for crossover gadgets,
which require the intersection square to be empty.
We fix these issues using shift gadgets.
This approach allows us to adjust the rows of the clause gadgets
and the columns of the clause propagators without parity restrictions.
If we put all the clause gadgets on rows of the same parity,
we can ensure that all the crossover gadgets line up.
We use the same trick to ensure proper chaining of the clause propagators
into the upper and lower AND gadgets,
and the chaining of those two AND gadgets into the final vertical AND gadget.
See Figure~\ref{entire_pic} for an example.

\section{Reduction Correctness}
\label{sec:correctness}

Finally we show that the construction of Section~\ref{sec:reduction}
is a valid reduction, i.e., an instance of RPM-3SAT is satisfiable
if and only the obtained ZHED puzzle is solvable.

\subsection{Satisfiable Formula $\Rightarrow$ Puzzle Solution}
\label{sec:intended}

Suppose there exists an assignment of the Boolean variables $v_i$s
that makes the formula evaluate to true.
Then we solve the constructed puzzle as follows.

First we expand each of the variable tiles.
If $v_i=\mathrm{true}$ in the solution,
we expand the corresponding $L$ tiles rightward;
otherwise, we expand the tiles leftward.
Then we expand all the vertical wires connecting variables to clauses,
toward the clauses.
Next we expand the threshold gadgets in the clause wires rightward.
Because each clause is satisfied in the formula,
each of the clause gadgets will be activated by at least one thick wire,
and hence each of the target squares for the clause gadgets
will be filled after this stage.

Now we expand all of the clause propagator wires.
These propagate the filled source corresponding to their clause's target square
to the extreme upper and lower horizontal AND gadgets,
filling all of their sources.
Next, we expand the upper and lower AND gadgets rightward,
which fills the two sources of the extreme-right vertical AND gadget.
Finally, we expand the last vertical AND gadget downwards,
which fills the puzzle target as desired.

\subsection{Puzzle Solution $\Rightarrow$ Satisfiable Formula}

Suppose we have a solution to the ZHED puzzle.
We argue about the different gadgets in the reverse order from the
intended activation sequence of Section~\ref{sec:intended},
and use the fact that a square can only be filled by gadgets
whose bounding boxes contain it, to obtain a satisfying Boolean assignment.
We call a threshold gadget \defn{successfully activated} if it fills its
target square.

\begin{itemize}
    \item The puzzle's target square only belongs to the bounding box of the extreme-right vertical AND gadget. Therefore that AND gadget was successfully activated, with at least $k=2$ of its sources filled by other gadgets.
    \item There are only $2$ sources of the extreme-right vertical AND gadget that might be filled by other gadgets, so both of them must have been. This implies that both the extreme upper and lower horizontal AND gadgets were successfully activated.
    \item For the extreme horizontal AND gadgets to be successfully activated, all clause propagators had to have successfully activated their target squares.
    \item This implies that all clause gadgets successfully activated their target squares.
    \item If none of the thick wires activated for any particular clause gadget, then there will not be enough sources in that clause's OR gadget that belong to other gadget's bounding boxes (specifically, crossovers) to reach its target square. Therefore, for every clause gadget, there is at least one thick wire with one or more component wires successfully activated.
    \item For any individual wire connecting a variable to a clause to be successfully activated, over half of the tiles in the corresponding variable gadget must be expanded in the direction of that clause (right if the clause is positive and left otherwise). We set each variable to false or true based on which direction (among left or right) the majority of the tiles of its corresponding variable gadget expanded; only clauses in this direction will be satisfied by the variable gadget. We set the value arbitrarily if there is a tie, in which case the variable did not satisfy any clauses.
    \item This assignment must be a satisfying assignment, because all clauses must have at least one variable satisfying them.
\end{itemize}

\subsection{Main Theorem}

We just established that determining whether a ZHED puzzle is solvable
is NP-hard.
It is also easy to prove membership in NP:
a solution can be described by a sequence of moves, and if there are $n$ tiles,
each move takes $O(\log n)$ bits to describe.
The number of moves cannot exceed the number $n$ of tiles,
so the total size of a proposed solution is polynomial in~$n$.
We can verify the solution in polynomial time by simply
maintaining the state of the board after each move and simulating the moves.
Together, these assertions prove our main result:

\begin{theorem}\label{Theorem:np-complete}
  It is NP-complete to decide whether a ZHED puzzle is solvable,
  even if all tile numbers are $1$.
\end{theorem}

\printbibliography

\end{document}